# INTERMEDIATE RESOLUTION SPECTROSCOPY OF THE RADIO GALAXY B2 0902+34 AT $z \approx 3.4$ [1]

J. M. Martín–Mirones, E. Martínez–González, J. I. González–Serrano & J. L. Sanz

Dpto. de Física Moderna, Universidad de Cantabria and
Instituto Mixto de Astrofísica y Estructura de la Materia, CSIC–Universidad de Cantabria
Facultad de Ciencias, Avda. de Los Castros s/n, 39005 Santander, Spain

---

[1] Based on observations made with the William Herschel Telescope

# ABSTRACT

We have carried out spectroscopic observations of the high redshift ($z \approx 3.4$) radio galaxy 0902+34 at intermediate resolution with the William Herschel Telescope. The dynamical spectral ranges covered are 4600 − 5480 Å and 5920 − 7680 Å with resolutions of 5.4 Å and 9.5 Å, respectively. We detect a continuum that is almost flat and also resolve three emission lines: Ly$\alpha$, C IV $\lambda1549$ and He II $\lambda1640$, the last one previously undetected. The line ratios are similar to the typical values found for narrow–line high redhift radio galaxies. Line ratios observed in different regions of the galaxy seem to indicate the presence of strong ionization and/or dust density gradients. We have not detected any Ly$\alpha$ absorption at $z = 3.3968$ (red wing of the Ly$\alpha$ emission line) as might be expected from the absorption found at 21 cm by other authors using the VLA and Arecibo antennas. We discuss possible models for the H I absorbing cloud.

*Subject Headings:* Galaxies: formation – Galaxies: individual: B2 0902+34 – Galaxies: structure

# I. INTRODUCTION

The study of high redshift objects is interesting to analyze the physical conditions of the early universe and the galaxy evolution at cosmological redshifts. In particular, the detection of tens of radio galaxies at $z > 1$ offer a good chance to carry out those studies (see McCarthy 1993 for a recent review). Most of these galaxies have been selected from low frequency, strong source radio surveys (primarily 3CR and 4C). The addition of a steep spectrum selection criterion has proven highly effective at selecting high redshift radio galaxies (Chambers et al. 1988b). Thus, the redshift record for galaxies, 4C 41.17 $z = 3.8$, has been obtained in the 4C Ultra Steep Spectrum by Chambers et al. (1990).

The optical identification of high redshift radio galaxies has shown that, in general, they are very extended and have relatively narrow permitted emission lines and strong forbidden ones, which distinguish them from QSOs. It has been claimed that some of these systems may be protogalaxies in the process of formation. Indications for this are the flat spectrum and the absence of the 4000 Å break, features which have already been observed in many cases. However, a fraction of the total light could come from nuclear activity. It is not clear whether the stellar emission dominates over the nuclear one. The discovery that a high fraction of high redshift radio galaxies show an alignment of their optical and radio emissions (McCarthy et al. 1987; Chambers et al. 1987, 1988a) suggests that there must be a connection between their optical and radio activities. This indicates a likely non–stellar origin of the optical emission.

The most common lines observed in these systems are Ly$\alpha$, C IV $\lambda1549$, He II $\lambda1640$, C III] $\lambda1909$, [O II] $\lambda3727$ and [O III] $\lambda5007$ (McCarthy 1993). Some of them could dominate the broad band emission producing in such cases an erroneous continuum spectrum, and therefore, overestimating the age of the galaxy. Data at wavelengths longer than 1 $\mu$m are required to determine the rest–frame optical properties of these high redshift galaxies (Lilly and Longair 1984; Eisenhardt and Lebofsky 1987), and hence to study their stellar populations.

We have carried out optical spectroscopic observations at intermediate spectral resolution of the massive high redshift radio galaxy 0902+34 at $z \approx 3.39$. This source belongs to the second Bologna survey (B2) with flux density $S_{408 \text{ MHz}} > 1$ Jy and was first identified by Lilly (1988) (from hereinafter L88). Observations in the spectral range from $V$ to $K$ suggest that it is a young galaxy (Eisenhardt and Dickinson 1992; Eales et al. 1993). Recent radio observations of the 21 cm line of neutral hydrogen have detected (Uson et al. 1991) and confirmed (Briggs et al. 1993) an absorption against the radio continuum source. This absorption could also leave a track in the optical, redwards the Ly$\alpha$ line. In this sense, Hippelein and Meisenheimer (1993) have detected a feature of this type studying the $z = 3.8$ 4C 41.17 radio galaxy and have interpreted it as a Ly$\alpha$ cloud (very common feature in the spectra of QSOs). One of the main motivations of our observation is the study of the possible Ly$\alpha$ absorption in 0902+34. Uson et al. (1991) also claimed to detect emission at $z = 3.3970 \pm 0.0003$ separated from the absorption by 33' but it has not been confirmed by recent radio observations (21 cm) by Briggs et al. (1993) using the Arecibo antenna and de Bruyn and Katgert (1993) using the Westerbork Radio Synthesis Telescope and by very recent narrow band Ly$\alpha$ observations by Martínez–González et al.

(1994).

In Section II, we describe the technical details of our observations. Section III includes the main results of the data analysis and their implications. Finally, in Section IV, we summarize the main conclusions of this work. The Hubble constant will be $H_0 = 50$ km s$^{-1}$ Mpc$^{-1}$ throughout this paper.

## II. OBSERVATIONS

We have carried out the spectroscopic observations at the 4.2 m William Herschel Telescope (WHT) at the observatory of Roque de Los Muchachos (Canary Islands, Spain) during the night of January 24, 1993. The seeing was of $\approx$ 1.2–1.6 arcsec. We have used the ISIS spectrograph on the Cassegrain focus of the WHT. In the blue and red arms of ISIS, we used the R600B and R316R gratings and the TEK1 and EEV3 CCD detectors, respectively. With this configuration, an intermediate spectral dispersion of 0.78 Å/pixel (blue arm) and 1.38 Å/pixel (red arm) is obtained. The spectral ranges are 4606–5482 Å for the blue arm and 5921–7681 Å for the red one. A long slit of width 3" was used providing a spectral resolution of $\approx$ 5.4 Å in the blue arm and of $\approx$ 9.5 Å in the red one. Both resolutions are a clear improvement over that achieved by L88 of 20 Å, allowing us to resolve the Ly$\alpha$ line (and its possible structure) and any other possible strong features appearing in the spectral range observed (e. g., C IV $\lambda$1549, He II $\lambda$1640, ..., the former already detected by L88). Given the redshift of 0902+34, the Ly$\alpha$ line would appear in the blue arm (close to the red end) and the other possible lines in the red one. Six different observations of 2700 s of the radio galaxy 0902+34 were carried out, all of them producing two subspectra corresponding to each arm of ISIS. We pointed to the radio core of the galaxy (09$^{\rm h}$ 02$^{\rm m}$ 24$^{\rm s}$.796, + 34° 19' 56".58, B1950.0) by performing offsets from nearby astrometric stars. The slit was rotated to coincide with the parallactic angle at the beginning of each exposure and is given in Table 1 (the slit orientation of the second exposure coincides with that used by L88 of 95°). This will allow us to map spectroscopically different regions of the galaxy. The reduction of the data was performed using the FIGARO package in the standard way. Finally, we obtained 12 sky–subtracted and flux–calibrated spectra.

## III. RESULTS AND DISCUSSION

We have made two separate analysis of the data: the first related to individual spectra corresponding to different extraction positions on the slit and the second to the summed spectrum. The latter spectrum would represent an average over a region 50° (area covered by individual exposures) around the Ly$\alpha$ axis of the radio galaxy (see the Ly$\alpha$ image given by Eisenhardt and Dickinson 1992). In the summed spectrum as well as in some individual ones, we have detected Ly$\alpha$, C IV $\lambda$1549 and He II $\lambda$1640, the last line detected for the first time in the present work. Also, a continuum emission is detected in the summed spectrum.

### a) Individual Spectra

In the analysis of the individual spectra, we have made Gaussian fits to the lines appearing in each spectrum. We do not detect continuum in any of the individual spectra and then we consider the continuum from the summed spectrum (fit obtained using the blue plus red arm points; see later) in the analysis of the individual ones. The noise levels used to calculate the reduced $\chi^2$'s come from 100 pixels in each side of the lines ($\approx 78$ Å in the blue arm and $\approx 138$ Å in the red one). Ly$\alpha$ has been detected and resolved in each of the six exposures. However, the C IV line is clearly detected in three of them, although it is resolved only in one. The He II line is detected and resolved in two exposures. In Table 1, we summarize the main results of these detections, including the best–fitted values of the three free Gaussian parameters (maximum intensity, mean wavelength and dispersion of the lines) and the corresponding reduced $\chi^2$, the redshift, FWHM, flux, equivalent width and luminosity of the lines and the C IV/Ly$\alpha$ and He II/Ly$\alpha$ flux ratios (the line fluxes in each ratio corresponding to the same exposure). We also show the extraction positions on the slit for each detection.

For the orientation of the slit along the Ly$\alpha$ axis (exposure 2 which coincides with the L88 P. A. of 95° and includes the two Ly$\alpha$ peaks shown in the image by Eisenhardt and Dickinson 1992), our results for the flux and luminosity are similar to those found by L88 for the central region of $3.5 \times 3.5$ arcsec$^2$. However, the equivalent width is less due to our larger value of the $V$ continuum by a factor $\approx 1.68$ (see later) and our less surface brightness by a factor $\approx 0.63$ (taking the size along the axis of the galaxy in the $V$ band to be 6 arcsec). Our better spectral resolution provides us with a slightly lower FWHM than that achieved by L88 (his small spectral resolution of 20 Å broadens the line). The height of the line is similar to that by L88.

For the other orientations of the slit, we have mapped fainter regions of the galaxy, that oriented at 269° being the most intense of them since this direction is almost the same as that with 95°. As can be seen from Table 1, a strong surface brightness gradient is present in 0902+34 since a little change in the orientation of the slit produces large changes in the fluxes, luminosities and equivalent widths. All these results are in agreement with the Ly$\alpha$ image of 0902+34 shown by Eisenhardt and Dickinson (1992), i. e., for off–axis orientations we map the region between the two peaks appearing in that image. From the redshifts listed in Table 1, the peaks of Ly$\alpha$ emission (present in exposures 2 and 5) seem to be slightly further than the remaining observed regions.

For C IV and He II, we see that the only common detection is that of exposure 2 where the lines are also resolved. Both lines are also detected in other orientations close to the axis but only He II can be resolved. The flux ratios seem to indicate a strong gradient in the degree of ionization of the gas being the two peaks of Ly$\alpha$ emission less ionized than the rest of the galaxy. On the other hand, from exposure 2 the FWHM of the C IV line is similar to that of the Ly$\alpha$ one, but the He II FWHM is approximately half of the latter. However, the redshifts of the three lines are approximately the same. These results seem to indicate that He II is emitted from a region of the galaxy different to that emitting Ly$\alpha$ and C IV. The distribution of redshifts might also imply that the C IV and the He II regions are more spread than the Ly$\alpha$ region, although the small signal present in the two former lines produce bigger errors than for Ly$\alpha$ (see Table 4 below).

### b) Summed Spectrum

We have also analyzed the spectrum obtained adding the sky–subtracted spectra corresponding to the six exposures (16200 s of integration) and flux–calibrating afterwards. In Figures 1 and 2, we show the blue and red parts of the spectrum filtered with a Gaussian FWHM equal to the resolution of the corresponding arm. From this spectrum, we have determined the continuum fitting the power–law $I_\lambda(\lambda) = K\lambda^\alpha$ to points obtained by averaging over regions free from known lines and sky features. So, we have used three points in the blue part within the wavelenght range 4700–5180 Å and five in the red one within the range 6400–7150 Å, excluding the ranges 6520–6610 Å (sky–line) and 6760–6850 Å (C IV line). The value of the continuum in each region is calculated by means of an iterative process. The fitted values are given in Table 2 and the corresponding power–law is plotted in Figures 1 and 2 (for comparison we also show the power-law resulting from the fit obtained using only the five points in the red part: $\log K (10^{-20}$ W m$^{-2}$ Å$^{-(1+\alpha)}$ arcsec$^{-2}$)= $6.65^{+1.60}_{-0.43}$, $\alpha = -2.3^{+0.1}_{-0.4}$). Moreover, in Table 2, we show the $V$ and $R$ (Kitt Peak broad–band filters) surface intensities and magnitudes.

Looking at Table 2, we see that the $I_\nu(\lambda)$ spectrum is almost flat ($I_\nu \propto \lambda^{-0.7}$) in the optical band in agreement with the less accurate results by L88. Although we find a continuum higher than that obtained by L88 ($\approx 40-70\%$ larger for the surface intensities, depending on the wavelength) both are compatible within the errors. The same is true for the $R$ data of Eisenhardt and Dickinson (1992). In any case, these results are compatible with the recent prediction (see, e. g., Eales et al. 1993 and Eisenhardt and Dickinson 1992) that 0902+34 is a young galaxy observed during its initial burst of star formation (the spectral energy distribution predicted for a galaxy in which there are no post–main–sequence stars is flat between the rest–frame wavelength of Ly$\alpha$ (1215.67 Å) and $\sim 5000$ Å).

From the summed spectrum we are able to resolve the three lines which appear in the individual spectra. In analysing these lines, we have tried to fit different models: a single emission (1E), two emissions (2E) and one emission and one absorption (1E1A), each component assumed to have a Gaussian profile. The first model have been used for the three lines whereas the other two only for Ly$\alpha$. Models 2E and 1E1A have been considered in order to study any possible structure present in the Ly$\alpha$ line. In particular, a possible absorption might be present as the optical counterpart of the 21 cm absorption found by Uson et al. (1991). Moreover, the presence of a possible Ly$\alpha$ absorbing cloud in front of the $z = 3.8$ 4C 41.17 radio galaxy has recently been claimed by Hippelein and Meisenheimer (1993). The continuum subtracted is that given in Table 2 and the noise level has been calculated as for the individual spectra. The reduced $\chi^2$'s found for the three Ly$\alpha$ fits are similar and less than 1 (see Table 3), and therefore they do not provide any evidence of structure in this line. From hereinafter we will only consider the simplest case (1E). For C IV and He II, we have not tried to study their structure because of their relatively small signal (the C IV emission is a doublet with two peaks at wavelengths 1548.188 Å and 1550.762 Å, the first one being twice more intense than the second). The significance of the detections of these two lines is $4.9\sigma$ and $2.9\sigma$ for C IV and He II respectively. Table 4 shows the physical characteristics of the lines in the summed spectrum. The high values of the errors quoted for some quantities are due to the propagation of the individual errors present in the fitted parameters of the lines assuming that these are independent random variables; therefore, they can be taken as upper and lower bounds better than as errors. In Figures 3, 4 and 5, we can see, in detail, the appearance of the three lines and the corresponding fits.

The redshifts of the radio galaxy 0902+34 obtained from the Ly$\alpha$ emission and the He II line are the same. However, the redshift obtained from the C IV line is compatible with the others only at $\approx 3\sigma$ level (the redshifts are determined from the mean wavelength of the Gaussian fits). Obviously, the redshifts obtained from the C IV and He II lines are less accurate due to the smaller signal relative to Ly$\alpha$. Also, the intrinsic doublet structure of C IV contributes to the inaccuracy in the determination of the redshift (in fact, to obtain the C IV redshift, we have used the weighted mean wavelength of the doublet: $\lambda = 1549.046$ Å). The values of the redshift obtained by us are slightly less ($\approx 3.391$) than those obtained in L88 of $z = 3.393$ (Ly$\alpha$) and $z = 3.399$ (C IV) and significatively more accurate due to our better spectral resolution.

The Ly$\alpha$ surface brightness is $\approx 3.2$ times less than the surface brightness for the central region obtained by L88 because of our surface brightness represents a mean over a wide angular area. This result is a consequence of the strong gradient in surface brightness present in this galaxy as was already mentioned when we discussed the individual spectra. In Table 4, surface luminosities for flat and open $\Omega = 0.1$ universes are also given. The line ratios obtained from the summed spectrum as well as from individual spectra (Tables 4 and 1) suggest that the ionization gradient decreases towards the Ly$\alpha$ axis of the galaxy. On the other hand, our Ly$\alpha$ equivalent width is less than that obtained by L88 (central region) by a factor $\approx 5.2$, coming from a factor $\approx 3.2$ in the surface brightness and $\approx 1.68$ in the $V$ continuum.

We have not detected any Ly$\alpha$ absorption as might be expected from the H I cloud observed in 21 cm by Uson et al. (1991) and Briggs et al. (1993) at $z = 3.3968 \pm 0.0001$ (redwards the Ly$\alpha$ emission) with column density $N_{\rm H\ I} \approx 10^{18} T_s$, where $T_s$ (°K) is the spin temperature. We can impose an upper limit to the maximum intensity of the absorption of $5.31 \times 10^{-4}$ mJy (assumed to have a Gaussian shape) based on the amplitude of the noise in a resolution element. Considering that the redshift of the absorbing cloud is higher than that of the Ly$\alpha$ emission a plausible model would be that the cloud is located in between the Ly$\alpha$ emitting region and the radio core. However, if the cloud would be placed nearer us than the Ly$\alpha$ emitting region (and therefore falling towards it with velocity $\approx 410$ km s$^{-1}$) then it would absorb a Ly$\alpha$ flux density of $S_{\rm Ly\alpha} = 3.3 T_s f$ (mJy) where $f$ is the filling factor. Taking into account the maximum absorption allowed by our data, we conclude that $f \leq 1.61 \times 10^{-6}$ if the spin temperature is as low as $T_s \approx 10^2$ °K (as found in nearby galaxies) and $f \leq 1.61 \times 10^{-8}$ if $T_s$ reaches $\approx 10^4$ °K as advocated by Zel'dovich (1970) and Sunyaev and Zel'dovich (1975) (these high temperatures are also derived from some quasar absorption systems). If we assume these two temperatures as extreme cases, the maximum angular size of the absorbing cloud would be in the range $\approx 0.00063 - 0.0063"$ giving a maximum linear size of $\approx 4.5 - 45$ pc for a flat universe and $\approx 9 - 90$ pc for a $\Omega = 0$ universe.

# IV. SUMMARY

From the intermediate resolution spectroscopic analysis of the radio galaxy 0902+34 in the optical band, we have arrived to the following conclusions:
- The Ly$\alpha$ ($\approx$ 660 km s$^{-1}$ FWHM) and C IV ($\approx$ 880 km s$^{-1}$ FWHM) lines have been resolved for the first time. Their surface brightnesses (W m$^{-2}$ arcsec$^{-2}$) and observed equivalent widths (Å) are respectively $\approx 2.9 \times 10^{-20}$ and $\approx 180$ for Ly$\alpha$ and $\approx 0.5 \times 10^{-20}$ and $\approx 55$ for C IV.
- The He II line ($\approx$ 540 km s$^{-1}$ FWHM) has been first detected and resolved in the present observation. It is characterized by a surface brightness of $\approx 0.3 \times 10^{-20}$ W m$^{-2}$ arcsec$^{-2}$ and an observed equivalent width of $\approx 40$ Å.
- The analysis of the line ratios in exposures with different orientations of the slit shows the existence of strong ionization and/or dust density gradients. Moreover, the ratios C IV/Ly$\alpha$ = 0.16, He II/Ly$\alpha$ = 0.11 and C IV/He II= 1.5 obtained from the summed spectrum are typical of radio galaxies with $z > 1.8$.
- We have also detected the optical continuum of the radio galaxy showing that it is almost flat in agreement with the recent conclusion, based on a much wider wavelength range (Eisenhardt and Dichinson 1992, Eales et al. 1993), that 0902+34 is a young galaxy observed during its initial burst of star formation. The higher continuum observed with respect to that of L88 together with the smaller line surface brightnesses decrease the rest–frame equivalent widths of the lines, resulting in values slightly smaller than the typical ones found for high redshift radio galaxies of $\gtrsim$ 100 km s$^{-1}$ for Ly$\alpha$ and $\gtrsim$ 50 km s$^{-1}$ for C IV and He II (McCarthy 1993).
- Finally, the possible Ly$\alpha$ absorption corresponding to the H I cloud observed in 21 cm at $z = 3.3968 \pm 0.0001$ has not been detected by us. This result implies that either the absorbing cloud is in between the Ly$\alpha$ emitting region and the radio core or it is placed nearer us than the Ly$\alpha$ emitting region but with a maximum angular size in the range $\approx 0.00063 - 0.0063$" (of the order of tens of parsecs) for spin temperatures in the range $\approx 10^4 - 10^2$ °K. The latter possibility is clearly more unlikely.

*Acknowledgments.* We acknowledge financial support from the spanish DGICYT, project PB92–0741. Partial finacial support for this project was provided by the Comission of the European Union and their Human Capital and Mobility Contract CHRX–CT92–0033. The William Herschel Telescope is operated by the Royal Greenwich Observatory at the spanish Observatorio del Roque de Los Muchachos of the Instituto de Astrofísica de Canarias, on behalf of the Science and Engineering Research Council of the United Kingdom and the Netherlands Organization for Scientific Research.

# FIGURE CAPTIONS

**Figure 1.** Summed (16200 s of integration) blue spectrum of 0902+34 filtered with a Gaussian FWHM equal to the resolution of the blue arm. We also show the best fits to the continuum obtained from all the points (3 points in the blue arm plus 5 points in the red one, solid line) and from only the 5 red arm points (dashed line) (see the text for a description of the procedure to obtain these points).

**Figure 2.** Same as Figure 1, but for the red spectrum. The feature present at $\approx$ 6300 Å is a sky–subtraction residual.

**Figure 3.** Portion of the summed blue spectrum containing the Ly$\alpha$ line. We also show the best–fit curve and the best–fit values of the parameters for the one emission model (1E).

**Figure 4.** Portion of the summed red spectrum containing the C IV line. We also show the best–fit curve and the best–fit values of the parameters for the one emission model (1E).

**Figure 5.** Same as Figure 4, but for the He II line.

# TABLE 1
# OBSERVED PROPERTIES OF THE LINES DETECTED IN THE INDIVIDUAL SPECTRA OF 0902+34

| Property | Lyα | | | | | | C IV | | | He II | |
|---|---|---|---|---|---|---|---|---|---|---|---|
| | Exp. 1 (82°)[a] | Exp. 2 (95°)[a] | Exp. 3 (128°)[a] | Exp. 4 (260°)[a] | Exp. 5 (269°)[a] | Exp. 6 (280°)[a] | Exp. 1[b] (82°)[a] | Exp. 2 (95°)[a] | Exp. 4[b] (260°)[a] | Exp. 2 (95°)[a] | Exp. 6 (280°)[a] |
| Maximum intensity ($10^{-20}$ W m$^{-2}$ Å$^{-1}$) | 3.250 | 6.250 | 4.500 | 3.250 | 6.000 | 2.750 | 0.950 | 0.475 | 0.825 | 0.875 | 0.975 |
| Mean wavelength (Å) | 5338.0 | 5338.6 | 5336.6 | 5336.2 | 5339.2 | 5338.6 | 6803.0 | 6799.0 | 6810.0 | 7201.0 | 7195.0 |
| Dispersion (Å) | 5.25 | 6.50 | 2.75 | 4.25 | 5.00 | 4.25 | 3.50 | 7.50 | 3.00 | 4.50 | 5.50 |
| $\chi_r^{2\,c}$ | 0.53 | 0.99 | 1.66 | 0.89 | 0.67 | 0.57 | 0.81 | 1.23 | 0.46 | 1.36 | 0.42 |
| Redshift | 3.3910 | 3.3915 | 3.3898 | 3.3895 | 3.3920 | 3.3915 | 3.3917[d] | 3.3892[d] | 3.3963[d] | 3.3909 | 3.3872 |
| FWHM (km s$^{-1}$) | 694.8 | 860.1 | 364.0 | 562.6 | 661.6 | 562.4 | 363.5 | 779.3 | 311.2 | 441.5 | 540.0 |
| Flux ($10^{-20}$ W m$^{-2}$) | 42.77 | 101.83 | 31.02 | 34.62 | 75.20 | 29.30 | 8.33 | 8.93 | 6.20 | 9.87 | 13.44 |
| Line flux/Lyα flux[e] | 1.00 | 1.00 | 1.00 | 1.00 | 1.00 | 1.00 | 0.19 | 0.09 | 0.18 | 0.10 | 0.46 |
| Equivalent width (Å) | 151.53 | 360.90 | 109.83 | 122.56 | 266.59 | 103.83 | 56.84 | 60.80 | 42.42 | 78.47 | 106.63 |
| Luminosity ($\Omega = 1.0$) ($10^{36}$ W)[f] | 3.88 | 9.25 | 2.81 | 3.14 | 6.83 | 2.66 | 0.76 | 0.81 | 0.57 | 0.90 | 1.22 |
| Luminosity ($\Omega = 0.1$) ($10^{36}$ W)[f] | 12.05 | 28.70 | 8.73 | 9.74 | 21.20 | 8.26 | 2.35 | 2.51 | 1.76 | 2.78 | 3.77 |

[a] Extraction angle on the slit ($\equiv$ parallactic angle at the beginning of the exposure).
[b] Unresolved line.
[c] Reduced $\chi^2$ of the best–fit to the line (the fitted parameters are the maximum intensity $I_{max}$, the mean wavelength $\langle\lambda\rangle$ and the dispersion $\sigma$).
[d] As rest–frame wavelength of C IV, we have adopted the weighted mean wavelength (1549.046 Å) obtained from the two peaks at 1548.188 Å and 1550.762 Å of the C IV doublet (the first being twice more intense than the second).
[e] The line flux and the Lyα flux used to calculate each quotient come from the same exposure.
[f] $H_0 = 50$ km s$^{-1}$ Mpc$^{-1}$. In each case, the redshift used is that for each line, i. e., both the regions emitting different lines and the pieces of the region emitting one line selected by the different orientations of the slit can be slightly displaced.

# TABLE 2
## CONTINUUM PROPERTIES OF 0902+34

| Property[a] | Value[b] |
|---|---|
| $\log K$ ($10^{-20}$ W m$^{-2}$ Å$^{-(1+\alpha)}$ arcsec$^{-2}$)[c] | $8.26^{+0.03}_{-0.75}$ |
| $\alpha$[c] | $-2.7^{+0.2}_{-0.0}$ |
| $V$ surface intensity ($10^{-20}$ W m$^{-2}$ Å$^{-1}$ arcsec$^{-2}$)[d] | $0.015^{+0.026}_{-0.012}$ |
| $V$ surface magnitude (mag arcsec$^{-2}$)[d] | $26.0^{+1.8}_{-1.1}$ |
| $R$ surface intensity ($10^{-20}$ W m$^{-2}$ Å$^{-1}$ arcsec$^{-2}$)[d] | $0.009^{+0.017}_{-0.008}$ |
| $R$ surface magnitude (mag arcsec$^{-2}$)[d] | $25.7^{+2.4}_{-1.2}$ |

[a] Assuming that the area of the galaxy covered in the $V$ band is 6"×3".
[b] Calculated from 8 points (3 blue arm points plus 5 red arm points).
[c] Power-law representation of $I_\lambda(\lambda)$: $I_\lambda(\lambda) = K\lambda^\alpha$.
[d] For the Kitt Peak broad-band interference filters whose central wavelengths are 5470 Å ($V$) and 6460 Å ($R$) and their bandpasses 940 Å ($V$) and 1260 Å ($R$).

# TABLE 3
## FITTED PARAMETERS OF THE LINES DETECTED IN THE SUMMED SPECTRA OF 0902+34

| Parameter[a] | Ly$\alpha$[b] | | | | | C IV[b] | He II[b] |
|---|---|---|---|---|---|---|---|
| | 1E | 2E | | 1E1A | | 1E | 1E |
| | | NE | BE | E | A | | |
| Maximum surface intensity ($10^{-21}$ W m$^{-2}$ Å$^{-1}$ arcsec$^{-2}$)[c] | $2.28^{+0.22}_{-0.17}$ | $1.81^{+0.42}_{-0.28}$ | $0.56^{+0.28}_{-0.28}$ | $2.36^{+0.14}_{-0.14}$ | $1.11^{+6.39}_{-0.42}$ | $0.21^{+0.06}_{-0.06}$ | $0.22^{+0.08}_{-0.07}$ |
| Mean wavelength (Å)[c] | $5337.9^{+0.5}_{-0.4}$ | $5338.0^{+0.6}_{-0.6}$ | $5337.0^{+5.5}_{-5.5}$ | $5337.4^{+0.5}_{-0.5}$ | $5332.6^{+0.3}_{-0.3}$ | $6811.0^{+11.5}_{-3.0}$ | $7201.0^{+2.5}_{-2.0}$ |
| Dispersion (Å)[c] | $5.00^{+0.50}_{-0.50}$ | $3.75^{+0.75}_{-0.50}$ | $11.00^{+8.00}_{-2.50}$ | $5.25^{+0.50}_{-0.25}$ | $0.70^{+0.20}_{-0.50}$ | $8.50^{+8.50}_{-2.00}$ | $5.50^{+2.50}_{-1.50}$ |
| $\chi^2_r$[d] | 0.84 | 0.82 | 0.82 | 0.80 | 0.80 | 0.82 | 0.79 |

[a] Assuming that the area of the galaxy covered in the V band is 6"×3".
[b] Models fitted: 1E ≡ one emission, 2E ≡ two emissions (NE ≡ narrow emission, BE ≡ broad emission), 1E1A ≡ one emission plus one absorption (E ≡ emission, A ≡ absorption).
[c] The errors are given by the 1-dimensional $1\sigma$ (68.3%) confidence contours.
[d] Reduced $\chi^2$ of the best-fit to the line (for the models 2E and 1E1A, $\chi^2_r$ is the reduced $\chi^2$ of the simultaneous fit to the two components).

## TABLE 4
## OBSERVED PROPERTIES OF THE LINES DETECTED IN THE SUMMED SPECTRA OF 0902+34 (MODEL 1E)

| Property[a] | Ly$\alpha$ | C IV | He II |
|---|---|---|---|
| Redshift | $3.3909^{+0.0004}_{-0.0003}$ | $3.3969^{+0.0074}_{-0.0019}$[b] | $3.3909^{+0.0015}_{-0.0012}$ |
| FWHM (km s$^{-1}$) | $662^{+66}_{-66}$ | $882^{+882}_{-209}$ | $540^{+245}_{-147}$ |
| Surface brightness ($10^{-20}$ W m$^{-2}$ arcsec$^{-2}$) | $2.85^{+0.56}_{-0.49}$ | $0.45^{+0.57}_{-0.22}$ | $0.30^{+0.25}_{-0.18}$ |
| Line surface brightness/Ly$\alpha$ surface brightness | 1.00 | 0.16 | 0.11 |
| Equivalent width (Å) | $182^{+36}_{-32}$ | $55^{+70}_{-28}$ | $43^{+36}_{-26}$ |
| Surface luminonity ($\Omega = 1.0$) ($10^{35}$ W arcsec$^{-2}$)[c] | $2.59^{+0.03}_{-0.03}$ | $0.41^{+0.43}_{-0.10}$ | $0.27^{+0.13}_{-0.08}$ |
| Surface luminosity ($\Omega = 0.1$) ($10^{35}$ W arcsec$^{-2}$)[c] | $8.04^{+0.08}_{-0.08}$ | $1.27^{+1.35}_{-0.32}$ | $0.84^{+0.39}_{-0.24}$ |

[a]Assuming that the area of the galaxy covered in the V band is 6"×3".

[b]As rest–frame wavelength of C IV, we have adopted the weighted mean wavelength (1549.046 Å) obtained from the two peaks at 1548.188 Å and 1550.762 Å of the C IV doublet (the first being twice more intense than the second).

[c]$H_0 = 50$ km s$^{-1}$ Mpc$^{-1}$. In each case, the redshift used is that for each line, i. e., the regions emitting different lines can be slightly displaced.